\documentclass{natureprintstylev4}

\usepackage{astjnlabbrev-nature}

\usepackage{amssymb}
\usepackage{mathptmx}
\usepackage{amsmath}	
\usepackage{gensymb}
\usepackage{color}
\usepackage{enumitem}

\usepackage{epsfig}
\usepackage{color}
\usepackage{graphicx}
\usepackage{longtable}
\usepackage{hyperref}

\usepackage[labelsep=endash]{caption}

\newcommand{\fast}{{\sc FAST}}

\newcommand{\msun}{${\rm M}_{\odot}$}

\newcommand{\msunyr}{${\rm M}_{\odot}~{\rm yr}^{-1}$}
\newcommand{\msunyrkpc}{${\rm M}_{\odot}~{\rm yr}^{-1}~{\rm kpc}^{-1}$}

\newcommand{\ohe}{\mbox{12\,+\,log(O/H)\,=\,}}
\newcommand{\ha}{{H$\alpha$}}

\newcommand{\angstrom}{\textup{\AA}}

\newcommand{\obj}{R5519}
 \newcommand{\zfourge}{{\sc zfourge}}                                                                                                                     
\newcommand{\citep}{\cite}
\newcommand{\citet}{\cite}

\title{A giant  galaxy in the young Universe with a massive ring}

\author{Tiantian~Yuan$^{1,2,*}$, Ahmed~Elagali$^{3,2}$, Ivo~Labb\'e$^{1}$, Glenn~G.~Kacprzak$^{1,2}$,  Claudia~del~P.~Lagos$^{3,2,4}$, Leo~Y.~Alcorn$^{5,6}$, Jonathan~H.~Cohn$^{5}$, Kim-Vy~H.~Tran$^{5,7,2}$,  
Karl~Glazebrook$^{1,2}$, Brent~A.~Groves$^{3,8,2}$, Kenneth~C.~Freeman$^{8}$, Lee~R.~Spitler$^{9,10,2}$, Caroline~M.~S.~Straatman$^{11}$, Deanne~B.~Fisher$^{1,2}$, Sarah~M. Sweet$^{1,2,12}$}

\begin{document}

\maketitle

\begin{affiliations}
\item Centre for Astrophysics and Supercomputing, Swinburne University of Technology, Hawthorn, Victoria 3122, Australia.
\item ARC Centre of Excellence for All Sky Astrophysics in 3 Dimensions (ASTRO 3D), Australia.
\item International Centre for Radio Astronomy Research (ICRAR), M468, The University of Western Australia, 35 Stirling Highway, Crawley, WA 6009, Australia.
\item Cosmic Dawn Center (DAWN), Denmark.
\item George P. and Cynthia Woods Mitchell Institute for Fundamental Physics and Astronomy, Department of Physics \& Astronomy, Texas A\&M University,
College Station, TX, 77843, USA.
\item Department of Physics and Astronomy, York University, 4700 Keele St., Toronto, Ontario, Canada, MJ3 1P3.
\item School of Physics, University of New South Wales, Sydney, NSW 2052, Australia.
\item Research School of Astronomy and Astrophysics, the Australian National University, Canberra, ACT 2611, Australia.
\item Research Centre for Astronomy, Astrophysics \& Astrophotonics, Macquarie University, Sydney, NSW 2109, Australia.
\item Department of Physics \& Astronomy, Macquarie University, Sydney, NSW 2109, Australia.
\item Sterrenkundig Observatorium, Universiteit Gent, Krijgslaan 281 S9, 9000 Gent, Belgium.
\item School of Mathematics and Physics, University of Queensland, Brisbane, Qld 4072, Australia
\end{affiliations}
\\
 
\begin{abstract}  %
In the local  ($z\approx0$) Universe, collisional ring galaxies make up only $\sim$0.01\% of galaxies\cite{Madore09} and are formed by  head-on galactic collisions that trigger radially propagating density waves\cite{Lynds76,Struck93,Appleton96}.
These striking systems provide key snapshots for dissecting galactic disks and are studied extensively in the local Universe\cite{Higdon95,Gerber96,Struck10,Mapelli12,Higdon15}.
However, not much is known about distant ($z>0.1$) collisional rings\cite{Lavery04,ElmegreenD06b,DOnghia08,Elagali18,Genzel14}.  
Here we present a detailed study of a ring galaxy at a look-back time of 10.8 Gyr ($z=2.19$).
Compared with our Milky Way, this  galaxy has a similar stellar mass, but has a stellar  half-light  radius that is 
 1.5$-$2.2 times larger and is forming stars 50 times  faster.   
The extended, diffuse stellar light outside of the star-forming ring, combined with a  radial velocity  on the ring  and  an intruder galaxy nearby, 
provides evidence for this galaxy hosting a collisional ring.  If the ring is secularly evolved\cite{Buta96,Comeron14}, the implied large bar in a giant  disk  would be inconsistent with the current  understanding  of  the earliest formation of barred spirals\cite{Sheth12,Kraljic12,Cen14,ElmegreenD14,Vincenzo19}.   Contrary to previous predictions\cite{Lavery04,DOnghia08,ElmegreenD06b}, this work suggests that massive collisional rings  were as rare 11 Gyr ago as they are today.  Our discovery offers a unique pathway for studying density waves in young galaxies,  as well as constraining the  cosmic evolution of spiral disks and galaxy groups.
\end{abstract}

The ring galaxy  (ID 5519, hereafter R5519)  was discovered in our systematic search for $z\gtrsim2$ spiral galaxies  in the   Cosmic Evolution Survey (COSMOS) field of the FourStar Galaxy Evolution survey (\zfourge\cite{Straatman16}).  We used  \zfourge\ catalogue images to identify spiral structures in galaxies  within the photometric redshift range of $1.8\lesssim z_{p} \lesssim 2.5$.   Owing to the surface brightness dimming and smaller sizes of galaxies at $z>1$, our visual  identification of spiral features was restricted to galaxies with illuminated pixels  larger than a radius  of $0.\!\!^{\prime\prime}$5 ($>$4 kpc at $z\approx2$) in the Hubble Space Telescope (HST)  images.  Our  visual inspection simultaneously identified ring galaxies and other morphologically distinct objects such as mergers and gravitationally lensed galaxies.  \obj\ was flagged as one of the largest galaxies among the $\sim$4000 galaxies inspected, with a clear ring structure as well as a large diffuse disk  (Fig.1 and Supplementary Figs.1-2).

We confirm the spectroscopic redshift of \obj\ to be   $z_s=2.192\pm0.001$ based on our Keck/MOSFIRE near-infrared (NIR) spectroscopy and  Keck/OSIRIS adaptive-optics aided NIR integral field spectroscopy (Supplementary Figs.3-4). A joint analysis of the MOSFIRE and OSIRIS spectroscopic data, in combination with the ground-based \ha\ narrow-band image from the \zfourge\ catalogue, shows that the \ha\ kinematics are consistent with a tilted rotating and expanding/contracting circular ring model (Fig.2, Methods). Taking the inclination angle ($i=29\degree \pm 5$) and the position angle ($PA=28\degree \pm10$) from an ellipse fit to the ring morphology (Supplementary Table 1, Supplementary Fig.1) as inputs to the kinematic model,  the inferred rotational velocity at the fixed ring radius  (R$_{ring}=5.1 \pm 0.4$ kpc) is V$_{\rm{rot}}$ {=} 90 $\pm$ 75 km/s and the radial expansion/contraction velocity is V$_{\rm{rad}}$ {=} 226 $\pm$ 90 km/s.  The velocity error bars represent uncertainties from observational measurements.  The systematic errors caused by  uncertain ranges of $PA$ and $i$ are of the same order of magnitude (Supplementary Figs.5-9). 

We verify that \obj\  resides  in a small  galaxy group environment, reminiscent of the loose groups in which local collisional ring galaxies (CRGs) such as the Cartwheel galaxy are found\cite{Higdon95,Appleton96,Madore09}. A companion galaxy (ID 5593, hereafter G5593) is confirmed  at a projected distance of $\sim$30 kpc from \obj\ with a 3D-HST survey\cite{Momcheva16} grism redshift of  $z_{grism}=2.184^{+0.005}_{-0.006}$  (Fig.1). An additional group candidate (ID 5475, hereafter G5475) is found at a projected distance of $\sim$40 kpc away (Fig.1), with a photometric redshift\cite{Straatman16} of $z_p=2.1\pm0.1$.  If G5593 is the intruder of \obj, then the inferred  timescale after collision is $\tau_{c}>$$39^{+65}_{-15}$ Myr; this is a lower limit due to unknown projection effects.
 
\obj\  has a total UV+IR star formation rate (SFR) of $80.0\pm0.2$ \msunyr and a stellar mass of  $\log({\rm M_*}/{\rm M}_{\odot}) =10.78\pm0.03$. In comparison, G5593 has a total SFR of $123\pm2$ \msunyr and $\log({\rm M_*}/{\rm M}_{\odot}) =10.40\pm0.04$ (Table 1, Methods). The furthest group candidate  (G5475)  within a 50 kpc projected distance from the ring is a  compact quiescent  galaxy. The morphology of G5593 shows double nuclei and a tidal tail (Fig.1), suggestive of an ongoing merger of its own. We find no active galactic nucleus (AGN) signatures in \obj\ nor in its companions based on  the \zfourge\ AGN catalogue\cite{Cowley16} and our MOSFIRE spectrum. The derived interstellar medium (ISM) properties indicate that \obj\ is metal-rich (\ohe8.6-8.8) (Supplementary Information, Supplementary Fig.3).

\obj's ring has the highest contrast in HST F125W and F160W  images (Fig.1). At $z_s=2.19$, F125W and F160W filters include contributions from strong emission lines such as  [OII]$\lambda$3727 and [OIII]$\lambda$5007, respectively.  The prominent ring structure in these bands is consistent with the ring being dominated by  emission from luminous star-formation regions. The  ground-based \ha\ narrow-band image suggests that the bulk  ($\sim$45-70\%) of  recent star formation occurs on the ring (Methods). We find tentative evidence for  the existence of an off-centre ``nucleus"  based on the redder colour  of one spatially resolved region in the deep  Ks$_{\rm tot}$ and HST near-infrared images (Methods, Supplementary Fig.2, Supplementary Table 2). For a range of extreme star-formation histories (10 Myr burst and constant SFR),  we derive for \obj\  a total stellar mass between 10$^{10.5}$ - 10$^{10.8}$ ${\rm M}_{\odot}$, and stellar ages between 0.05-2 Gyr (Methods, Supplementary Fig.10).

The ring of \obj\ is small compared with local CRGs.  For example, 90\%  of  rings in the local CRG sample\cite{Madore09} have ring radius $\gtrsim$5 kpc.  \obj's SFR  is at least 4$\times$ larger than local CRGs\cite{Romano08} of similar stellar masses (Table 1, Supplementary Fig.11). The enhanced SFR of \obj\ compared with local CRGs is understandable in the context that high-$z$ star-forming galaxies  have a  larger  molecular gas fraction\cite{Tacconi10}.   The average star formation rate surface density ($\Sigma$$_{\rm SFR}$) of \obj\ is $\sim$0.3~\msunyrkpc, typical of a star-forming galaxy at $z\approx2$, and   $\sim$4-8 times larger than local CRGs such as  Arp 147 and the Cartwheel\cite{Fogarty11,Higdon15}.  Interestingly, nearby CRGs  show moderately elevated SFR  relative to $z\sim0$ isolated disks\cite{Romano08}, whereas  \obj\  does  not have a substantially higher SFR in relation to its $z\approx2$  peers (Supplementary Fig.11). Both \obj\  and its companion G5593 lie within the 0.3 dex scatter of the ${\rm M_*}$-SFR ``main-sequence"\cite{Pearson18} relation of star-forming galaxies at $z\approx2$.  CRGs  are rare laboratories to study star formation in interacting galaxies\cite{Appleton96,Higdon15}. Future observations on the molecular gas would be important in revealing the details of the star formation processes in  \obj.

One of the most striking features of \obj\ is the extended stellar light outside of the ring  in multiple wavelengths (Fig.1, Supplementary Fig.2).  We have ruled out \obj\ as a regular merger or a gravitationally lensed system (see  Supplementary Information). We quantify the size of the diffuse light by measuring  R$_{80}$, the radius within which  80\%  of the total luminosity is included (Methods, Table 1, Supplementary Figs.12-13).  Comparing with  other  $z \sim 2$ galaxies  in the 3D-HST catalogue\cite{vander14},  \obj's ${\rm R}_{80}$ is  2.4$\sigma$ larger than  the mean size (5.4 kpc) of all late-type galaxies ($\log({\rm M_*}/{\rm M}_{\odot})>9.5$) and  1.5$\sigma$ larger than the mean value (7.1 kpc) of the most massive  ($\log({\rm M_*}/{\rm M}_{\odot})\ge10.6$) late-type galaxies at $z\approx2$ (Fig.3).  A morphological inspection on the other unusually large ($1\sigma$ above the mean) and massive late-type galaxies in the COSMOS field reveals that  most (4/7) of them are probably mergers (Fig.3).  Excluding the 4 mergers, the mean value of the most massive galaxies at $z\approx2$ is 6.4 kpc and is 2.6$\sigma$ lower than \obj.  Compared with our Milky Way's stellar  disk, the half-light radius of \obj\ is 1.5-2.2 times larger and its  ${\rm R}_{80}$ is 1.2-1.8 times larger (Methods).
 
If \obj\ is a secularly evolved resonant ring\cite{Buta96} (see Supplementary Information), then the giant disk and the implied large bar (half length $\sim$5 kpc, similar to the Milky Way's bar)   is challenging to understand at this redshift.  Diffuse stellar disks and/or bars as large as those of \obj\ have  not been conclusively reported in observations or simulations at $z>2$ [ref.\cite{Genzel14,Sheth12,Kraljic12,ElmegreenD14}].   For the rare, smaller ($<$1 kpc in radius)  barred spiral galaxies formed in simulations at $z>2$ [ref.\cite{Sheth12,Cen14,Vincenzo19}],  they are relatively isolated and do not reside in an active environment like \obj. Our interpretation of a collisional ring instead of a secularly evolved ring  can be verified by high spatial resolution imaging with the {\it James Webb Space Telescope} (JWST) in the mid-IR wavelength.

If \obj\ is exhibiting a first ring after collision in the classic model of an expanding wave\cite{Appleton96,Gerber96,Mapelli12}, the large R$_{disk}$ and small R$_{ring}$ imply  different  collisional timescales ($\tau_{c}$ $\ge$ 80 Myr and $\tau_{c}$ $\le$ 50 Myr, see Supplementary Information). The  inconsistency in  $\tau_{c}$  can be reconciled if the current ring in \obj\ is a second  ring after the collision.  A local analogy would be the Cartwheel galaxy\cite{Higdon95,Higdon15}, when Cartwheel's  outer ring fades and the inner second ring dominates. A second ring would explain the small ratio of R$_{ring}$ and R$_{disk}$. The large diffuse emission can  be accounted for as the first expanding ring sweeps up the pre-collisional disk. The thickness (3.7$\pm$0.3 kpc; Methods) of the ring,  the size ratio of the first to the second ring ($\sim$2.2, taking R$_{80}$ in the rest-frame optical as the radius of the first ring), SFR, age, and metallicity are broadly consistent with  the analytical model predictions of successive rings\cite{Struck10}.

The diffuse light induced by the expanding ring is difficult to observe in local CRGs because of the low surface density of the redistributed stars\cite{Romano08}.  \obj's extended disk has a rest-frame B-band (HST F160W) surface brightness of   $\sim$20 AB mag~arcsec$^{-2}$. Such a bright outer disk has yet to be  seen in local CRGs (Supplementary Fig.14). The diffuse emission outside of \obj's ring ($\gtrsim6.5$ kpc) contains $\gtrsim$50\% of the total light in the rest-frame B band, whereas for local CRGs, most of the B-band luminosity is  on and within the ring (Supplementary Fig.14). Without an intrinsic luminosity evolution with redshift, local CRGs' rings and extended disks would be undetectable  at $z\approx2$  with current observations (see Supplementary Information).  However, if high-redshift CRGs follow the ``main-sequence" relation at $z\approx2$ as R5519 does, then they  would  be bright enough to be detected in the 3D-HST WFC3 images (Supplementary Fig.14).

An alternative scenario  to explain the large diffuse light outside of the ring is through satellite perturbations. Recent CRGs  in the Evolution and Assembly of GaLaxies and their Environments (EAGLE) simulations\cite{Elagali18} show that  interaction with multiple satellites at $z>2$ can cause $>$50\% of the stellar particles of the CRG host to be tidally perturbed outside of the ring   $\sim$120 Myr after the collision (Supplementary Figs.15-16). Similar to the EAGLE ring, the diffuse stellar light of \obj\ could be tidally induced by small satellites or represents  an ongoing accretion of small satellites. In this scenario,  the ring can be either  the first or a successive ring.

Both our observation,  and  the EAGLE simulations, imply that the volume number density of massive ($\log({\rm M_*}/{\rm M}_{\odot}) \gtrsim 10.0$) CRGs at $z\approx2$  is as small as $z\sim0$ (Supplementary Information).  This seems contrary to previous predictions that CRGs are more common at high redshift\cite{Lavery04,ElmegreenD06b,DOnghia08}.  Using a scaling relation of $(1+z)^{4.5}$ from a previous study\cite{Lavery04}, CRGs are expected to be $\sim$140 times more common at $z\approx2$. Considering only the massive CRG hosts, the expected number density at $z\approx2$ is still  $\gtrsim$10$\times$ larger than  at $z\sim0$ (Methods).  We speculate  that  a combined effect of a rising merger rate, a decreased fraction of large spiral disks, and the lack of local-like galaxy groups  at high redshift could cause the slow CRG number density change\cite{Elagali18} in the past 11 Gyr (Supplementary Information).  If  \obj\ is a density wave ring similar to local CRGs,  it  is an unequivocal sign of the existence of a thin disk in the young universe, critical for understanding the onset of spiral galaxies\cite{ElmegreenD14,Yuan17,Vincenzo19}.

\begin{flushleft}
{\bf \large References}
\end{flushleft}

\begin{addendum} 

\item[Author Information] 
Correspondence and requests for materials should be addressed to T.Y. (tiantianyuan@swin.edu.au).

\item[Acknowledgements] 
This research was supported by the Australian Research Council Centre of Excellence for All Sky Astrophysics in 3 Dimensions (ASTRO 3D), through project number CE170100013. 

\item[Author Contributions] 
T.Y.  wrote the manuscript and had the overall lead of the project.  
A.E. carried out the EAGLE simulation analysis and contributed to the writing of the simulation results. 
I.L.,G.K. and C.L. contributed significantly to the overall science interpretation,  data analysis and making of the figures. 
L.A.,J.C., K-V.T., K.G. contributed significantly to the photometric and kinematic data analysis.
All other coauthors contributed by their varied contributions to the science interpretation, data analysis and
assistance with Keck observations.  All co-authors contribute to the commenting on this manuscript as part of an internal review process. 

\item[Competing interests statement] The authors declare no competing interests.

\end{addendum}

\begin{figure*}[!ht]
\centerline{
\includegraphics[width=0.9\textwidth, trim=0.25cm 0.25cm 0.25cm 0.5cm,clip=true]{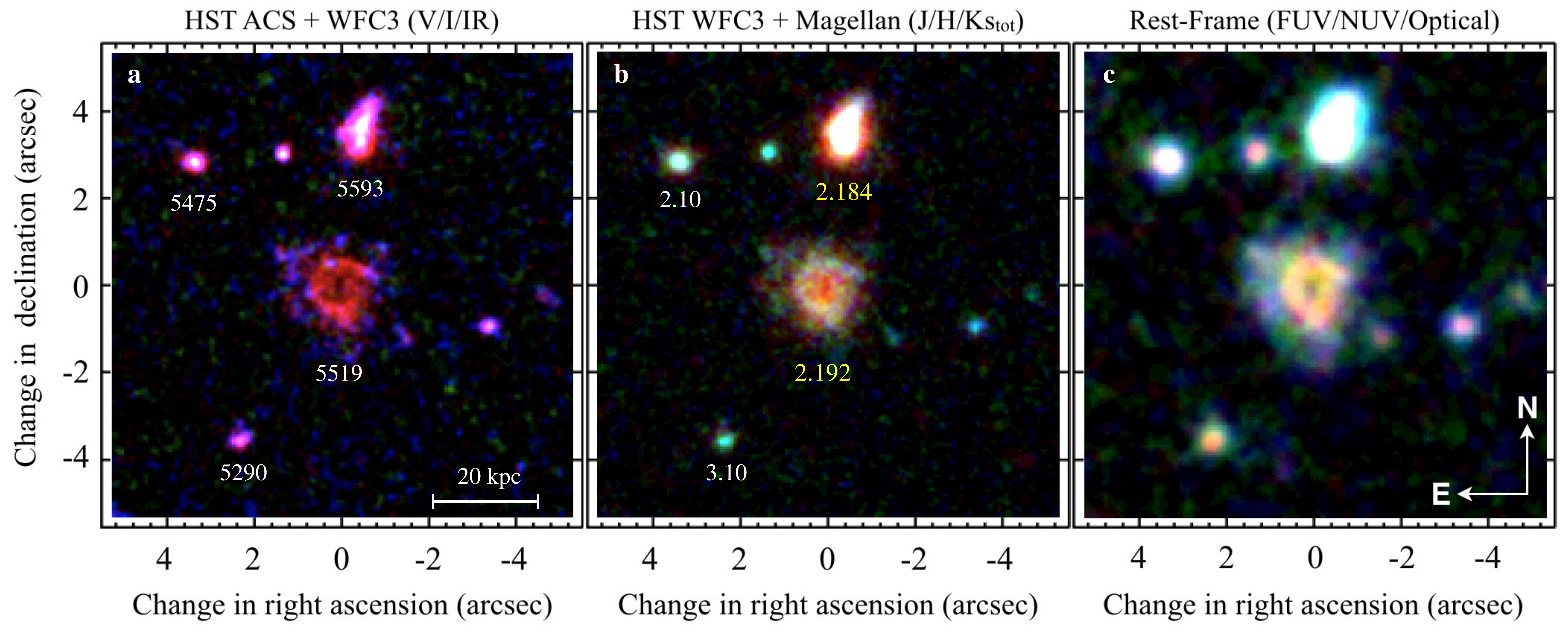}
}
\caption{\textbf{Multi-wavelength views of \obj\ and and its neighbouring environment}.  
{\bf a,}  a three-colour image combining HST V-band/ACS F606W (blue), I-band/ACS F814W (green), and IR-band/WFC3 F125W+F140W+F160W (red) images. 
Panel {\bf a} shows the 
highest spatial resolution (full width at half maximum $\sim$1.7-2.2 kpc) view of  \obj. The numbers under each object denote the \zfourge\ catalogue ID. 
The contrast of the image is tuned to highlight the  double nuclei  and tail-like structure of the companion galaxy  G5593. 
{\bf b,} Blue/Green/Red (HST F125W/F160W/Magellan Ks$_{\rm tot}$) colour image. 
The Ks$_{\rm tot}$ image is a super deep K-band detection image in \zfourge\ (Methods).
This image
highlights the resolved ring structure on top of the longest-wavelength image from the ground-based \zfourge\ Ks$_{\rm tot}$ band (rest-frame R). 
The \zfourge\ photometric redshifts are labelled in white, with confirmed  spectroscopic redshifts in yellow.   
{\bf c,} Blue/Green/Red (combined rest-frame FUV/NUV/optical) colour image. 
The images are generated by stacking  HST and  \zfourge\ catalogue images that correspond to rest-frame FUV, NUV, and optical wavelengths (see Supplementary Information). 
The pixel scale for image {\bf a} is  $0.\!\!^{\prime\prime}06$ as sourced from the 3D-HST survey. For {\bf b-c},   a pixel scale of $0.\!\!^{\prime\prime}15$ is used to 
match the ground-based images\cite{Straatman16}. A logarithmic stretch is used for all images in this work.
 \label{fig:fig1}
}
\end{figure*}

\begin{figure*}[!ht]
\centerline{
\includegraphics[width=0.9\textwidth, trim=0.5cm 0.1cm 1cm 0.1cm,clip=true]{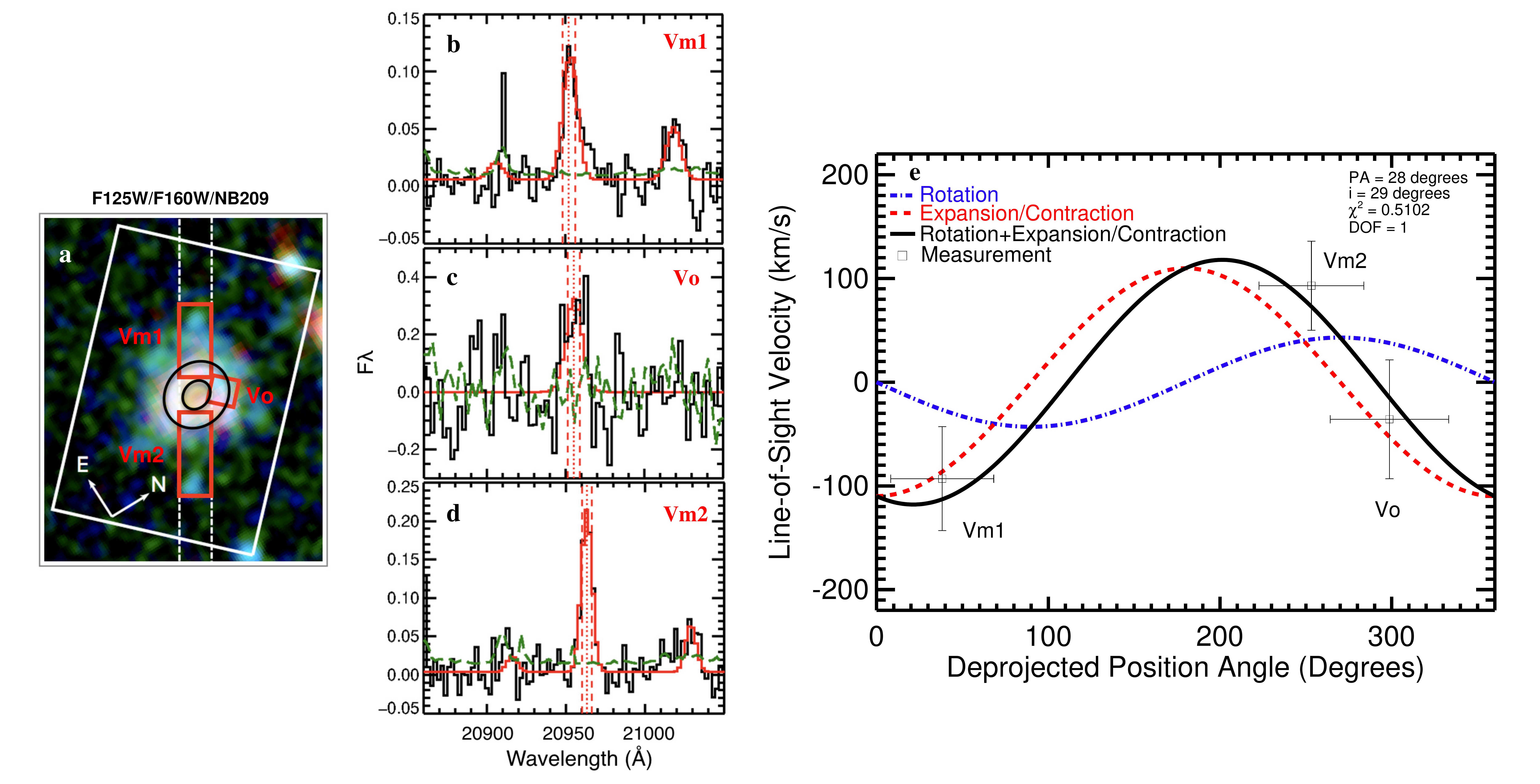}
}
\caption{\textbf{A joint analysis of the MOSFIRE and OSIRIS ring kinematics}.  
{\bf a,} Spatial alignment of the OSIRIS (white box; field-of-view  $4.\!\!^{\prime\prime}5\times 6.\!\!^{\prime\prime}4$) and MOSFIRE (white dashed  box; slit width $0.\!\!^{\prime\prime}8$) observations
on the Blue/Green/Red three-colour image (HST F125W/F160W/ \zfourge\ NB209 \ha\ narrow band).  The black circles show the best-fit double ellipse to the ring structure.
 The red boxes mark the three position angles where the line-of-sight velocities  (MOSFIRE: V${\rm{m1}}$, V${\rm{m2}}$;  OSIRIS: V${\rm{o}}$) are measured (see also Supplementary Figs.5-6). 
{\bf b-d,}  The observed \ha\ spectra  (black), best-fit (red)  and 1$\sigma$ noise (green) for V${\rm{m1}}$, V${\rm{m2}}$ and V${\rm{o}}$.  
F$\lambda$ is the observed flux of the spectra.
The red vertical dotted and dashed  lines show the the determined velocity centre and its uncertainty range, respectively. 
The uncertainty range  is a conservative measure using the 1$\sigma$ width of a Gaussian line profile (Methods). 
The flux unit for {\bf b, d} is 10$^{-17}$ergs/s/cm$^{2}$/$\angstrom$, and for {\bf c}, the flux unit is 10$^{-20}$ ergs/s/cm$^{2}$/$\angstrom$. 
{\bf e,} Fitting an  expanding/contracting tilted circular ring model to the line-of-sight velocity versus deprojected position angle diagram.
The  vertical errors  of V${\rm{m1}}$, V${\rm{m2}}$, and  V${\rm{o}}$ are defined by the uncertainty range of the line centre in {\bf b-d}; the horizontal error 
is defined by the alignment uncertainty in  {\bf a}.  The model uses a fixed inclination  $i=29\degree$ and a  kinematic major axis of $PA=28\degree$.  
  The black curve is the best-fit  model with  contributions from both rotation (blue) and expansion/contraction (red).   The expansion/contraction component is  detected for a varied range of $PA=0-45\degree$ and $i=20-45\degree$ (Methods, Supplementary Figs.7-9).
   \label{fig:kine}}
\end{figure*}

\begin{figure*}[!ht]
\centerline{
\includegraphics[width=0.8\textwidth, trim=0.4cm 0.25cm 0.25cm 0.25cm,angle=0,clip=true]{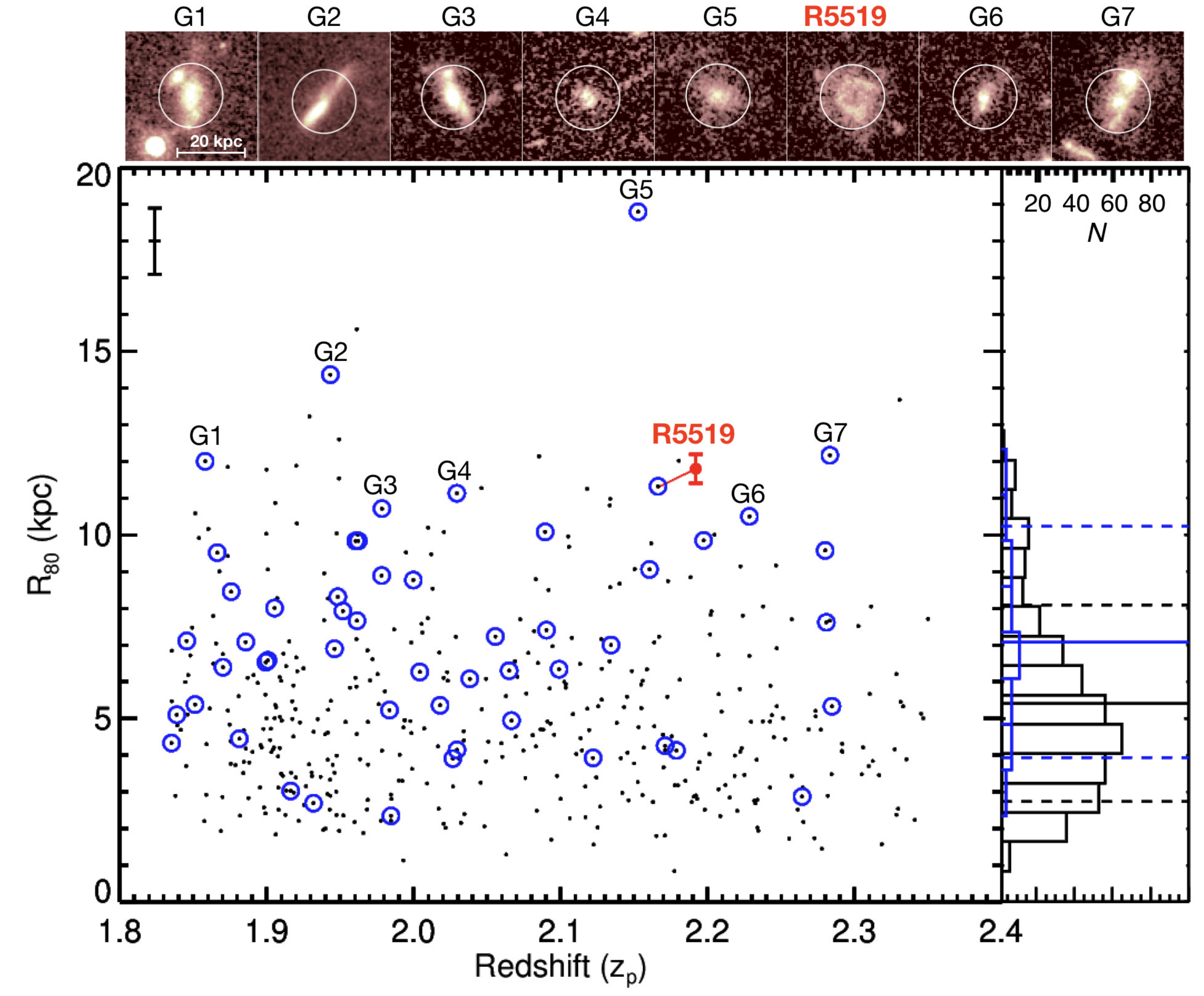}
}
\caption{\textbf{Comparing the size of \obj\ with the size distribution of late-type galaxies at $z\approx2$}.  
The $z_{\rm p}$ and R$_{80}$   values are from the 3D-HST  survey  and its mass-size catalogue in the COSMOS field (Methods). 
The typical error bar for  $\rm{R}_{80}$, defined as an average  error propagated through  the 1$\sigma$ model error of the effective radius in 3D-HST,  is shown in the top left corner.  
The black dots show all galaxies with  $\log({\rm M_*}/{\rm M}_{\odot})\ge9.5$. 
The blue circles highlight the most massive objects, defined as $\log({\rm M_*}/{\rm M}_{\odot})\ge10.6$. 
The histograms (right) show the size distribution for all (black) and the most  massive (blue) galaxies. 
The horizontal solid and dashed lines mark the mean and the 1$\sigma$ scatter of the size distributions. 
 Including \obj, there are eight objects (labelled G1-G8) that have unusually large sizes (defined as  1$\sigma$ above the mean size of the most massive galaxies). 
 The F160W band morphologies for these eight objects are shown as  postage-stamp images at the top.  
The  red filled circle  shows  $z_{\rm s}$ and our non-parametric measurement of R$_{80}$ for \obj. We note that the inferred 
 R$_{80}$ for the 3D-HST sample is based on an empirical relation for a large sample of  galaxies;  individual galaxies may deviate from this relation due to scatter  (Methods). 
\label{fig:fig3}} 
\end{figure*}

\begin{table*}[!ht]
\setlength{\tabcolsep}{0.4em}
\renewcommand\arraystretch{0.2}
\begin{center}
\begin{tabular}{lccc}
\hline
\hline
Objects   & \obj\ &  G5593   & G5475     \\[0.92ex]
 \hline\hline
Right ascension (J2000)   & 10$^{h}$00$^{min}$43.847$^{s}$ & 10$^{h}$00$^{min}$43.815$^{s}$ &10$^{h}$00$^{min}$44.081$^{s}$  \\[0.92ex]
\hline
Declination (J2000)    & +2$^{\degree}$14$^{\prime}$ 05.476$^{\prime\prime}$  &+2$^{\degree}$14$^{\prime}$ 09.167$^{\prime\prime}$ & +2$^{\degree}$14$^{\prime}$ 08.469$^{\prime\prime}$ \\[0.92ex]
\hline
Redshift $z$ & 2.1927$^{+0.0001}_{-0.0001}$ &  $2.184^{+0.005}_{-0.006}$ & 2.10$^{+0.07}_{-0.07}$  \\[0.92ex]
 \hline
 Stellar mass   (log$_{10}$(M$_*$/M$_{\odot}$))  & 10.78$^{+0.03}_{-0.03}$  &10.40$^{+0.04}_{-0.04}$&9.55$^{+0.03}_{-0.04}$ \\[0.92ex]
 \hline
 SFR$_{\rm IR+UV}$  (M$_{\odot}$ yr$^{-1}$)   & 80$^{+0.2}_{-0.2}$&123$^{+2}_{-2}$&0 \\[0.92ex]
 SFR$_{\rm H\alpha}$  (M$_{\odot}$ yr$^{-1}$)   &38-48& -& -\\[0.92ex]
  \hline
  &0.2-0.4& $\sim$0.54& - \\[0.92ex]
$\Sigma$$_{\rm SFR}$ (M$_{\odot}$ yr$^{-1}$ kpc$^{-2}$)   &&& \\[0.92ex]
  \hline
   $i$ (${\degree}$) & $29^{+5}_{-5}$ &-& -  \\[0.92ex]
  \hline
 $PA$ (${\degree}$) & 28$^{+10}_{-10}$ &-& -  \\[0.92ex]
  \hline
   Ring average radius  (kpc) & 5.1$^{+0.4}_{-0.4}$ & - & -  \\[0.92ex]
 ~~~~~~~~~~~~ inner radius (kpc)  & 2.7$^{+0.2}_{-0.2}$ & - & - \\[0.89ex]
 ~~~~~~~~~~~~ outer radius  (kpc) &   6.5$^{+0.2}_{-0.2}$& - & -  \\[0.89ex]
  ~~~~~~~~~~~~ ring thickness   (kpc) &   3.7$^{+0.3}_{-0.3}$& - & -  \\[0.89ex]
 \hline
  R$_{50}$ (kpc)    & 7.6$^{+0.2}_{-0.2}$ & 3.3$^{+0.1}_{-0.1}$ &  1.5$^{+0.2}_{-0.2}$  \\[0.92ex] 
  \hline
 R$_{80}$ (kpc)   & 11.8$^{+0.3}_{-0.3}$ & 4.9$^{+0.2}_{-0.2}$ & 2.5$^{+0.2}_{-0.2}$   \\[0.92ex]
  \hline
   Rotational velocity V$_{\rm{rot}}$  (km~s$^{-1}$) & 90$^{+75}_{-75}$ &-& - \\[0.92ex]
  \hline
   Radial (expansion) velocity V$_{\rm{rad}}$  (km~s$^{-1}$)  & 226$^{+90}_{-90}$ &-& - \\[0.92ex]
\hline\hline  
\end{tabular}
\end{center}
\caption{{\bf Physical properties of \obj\ and its neighbours.} \small The $z$ values for \obj, G5593, and G5475 are  MOSFIRE $z_{\rm s}$, HST $z_{\rm grism}$, and \zfourge\ $z_{\rm p}$, respectively.  For all sources, the  stellar masses are calculated based on the spectral energy distribution from the \zfourge\ photometry and 
the SFR$_{\rm IR+UV}$ are derived on the basis of the combined rest-frame IR and UV luminosities.
The range of SFR$_{\rm H\alpha}$ values reflects different assumptions about attenuation laws.   $\Sigma$$_{\rm SFR}$ is estimated by dividing  
the total SFR by an assumed total SFR area, with the range  of values reflecting  different assumptions (Methods).   
PA and $i$ are measured on the basis of the ellipse fit to the combined HST F125W+F140W+F160W image.   
The inner and outer radii and  thickness of the ring are derived from a double ellipse fit, whereas the average radius is based on a single ellipse fit (Methods,Supplementary Fig.1).  
R$_{50}$ and  R$_{80}$ are the radii where  50\% and 80\% of the total luminosity are enclosed in the combined HST F125+F140W+F160W band, respectively.  
The rotational and radial velocities are derived at the 5 kpc ring radius  using a tilted rotating and expanding circular ring model (Methods). 
}
\label{tab:tab1}
\end{table*}

\clearpage

\methods
Throughout we adopt a  $\Lambda$ cold dark matter cosmology, where $\Lambda$ is the cosmological constant, 
with $\Omega_{M}{=}0.307$, $\Omega_{\Lambda}{=}0.693$, and H$_{0}{=}67.7$~km~s$^{-1}$~Mpc$^{-1}$, consistent with the Planck measurements\cite{Planck14} and the cosmological parameters used in EAGLE simulations\cite{Elagali18}.  At the redshift of  $z=2.19$,  the look-back time is 10.8 Gyr and one arcsecond corresponds to a physical scale of 8.49 kpc.  All magnitudes are in AB units\cite{Oke83}, unless otherwise specified. 

\section {Size of the Ring and the Extended Diffuse Emission}\label{sec:sz}
\subsection{Size of the ring}\label{sec:ringsize}
We use the original 3D-HST CANDELS version of the F125W, F140W and F160W images\cite{Grogin11,Skelton14} to quantify the size of the ring structure. We fit both a single ellipse and a double ellipse of a constant width to the imaging data using a $\chi$$^{2}$ minimisation procedure.  In both approaches, we divide the ring azimuthally into N=180 intervals and use the average full width at half maximum (FWHM)  of the surface brightness (SB) along each azimuthal angle interval to determine the inner and outer edges of the ring. The baseline of the FWHM is chosen as the average SB in the central pixels of the ring (red cross in Supplementary Fig.1). We smooth the images by 3 pixels ($0.\!\!^{\prime\prime}18$) to enhance the signal-to-noise  ratio (SNR).

In the single-ellipse approach, we use data points on the outer edge of the ring as the input   and weight each azimuthal angle with the median SB within each azimuthal interval. The free parameters of a single ellipse model are:  the centre ($xc, yc$), major axis radius ($a$), axis ratio ($b/a$ or inclination angle $i$) and position angle (PA).  In the double ellipse approach, we use the inner and  outer edges of the ring as double constraints and weigh each azimuthal angle  with the median SB within each azimuthal interval.  The additional free parameter in the  double ellipse model is the width ($\Delta$R) of the ring. We find that both approaches provide reasonably good fits to the data (Supplementary Table 1).  We carry out the fit on both the single bands (F125W, F160W) and the combined band (F125W+F140W+F160W).   The results  are consistent within  1$\sigma$ of the  statistical errors.  The best-fit ellipses and parameters are shown in Supplementary Fig.1 and Supplementary Table 1.

\subsection{Size of the diffuse stellar light}\label{sec:diffuselight}
We quantify the size of the diffuse stellar light in a model-independent way by measuring the accumulated luminosity within a circular aperture of an increasing radius (Supplementary Fig.12). We calculate R$_{50}$,  R$_{80}$, and R$_{95}$  where   50\%,  80\% and 95\% of the total luminosity is enclosed. The choice of the three radii is to facilitate comparison with other samples of galaxies:  R$_{50}$ is comparable to the effective radius R$_{e}$ of a Sersic profile; R$_{80}$ is a recently popularised  parameter   to study the size evolution of  galaxies with redshifts\cite{Miller19}; R$_{95}$  describes the outer edge of the galaxy.  For a Sersic profile of $n=0.5$ and $n=1.0$,   R$_{95}$ corresponds to a radius of $\sim$2.1R$_{e}$ and $\sim$2.9R$_{e}$, respectively\cite{Graham05}.

The diffuse stellar light of \obj\ is  present in multiple wavebands (Supplementary Fig.2).   To test the dependence of the measured size on wavelength, image depth, and the  point spread function (PSF), we carry out the measurement on the deep  \zfourge\ Ks$_{\rm tot}$ image and the high-spatial resolution HST F125+F140W+F160W image; we then  repeat the measurements on the PSF matched \zfourge\ image, including a stacked rest-frame FUV image.  The PSF is best characterised by a Moffat profile of two parameters: FWHM and $\beta$, where $\beta$ describes the overall shape of the PSF.
The PSF matched images are carefully generated by the \zfourge\ team using  a Moffat profile with FWHM of $0.\!\!^{\prime\prime}9$ and $\beta=0.9$ [ref.\cite{Straatman16}]. Our stacked rest-frame FUV image is produced from the PSF-matched ground-based UBGV images (see Supplementary Information).  

We summarise the  derived R$_{50}$,  R$_{80}$, and R$_{95}$  in Supplementary Table 4.  The error bars are derived by perturbing the measurements within $1\sigma$ of the sky background.  The PSF-matched images yield on average a larger size of 1.4$\pm$0.6 kpc  at all wavelengths.   For images of similar depth and PSF, the bluest wavelength yields the largest size, e.g., the rest-frame FUV size is $\sim$1 kpc larger than the rest-frame B band.  Note that the diffuse stellar light distribution of \obj\ is not circularly symmetric, our R$_{50}$,  R$_{80}$, and R$_{95}$    can be considered as  circularly averaged  values. These circularly averaged values are consistent with the size measurement from the surface brightness distribution along the  major axis of the ring ellipse below.

\subsection{Surface brightness profile in 1D along the major axis}\label{sec:sb}
We measure the  1D surface brightness distribution  SB(\rm{R}) by averaging three slices along the major axis of the ring ellipse in the deep HST F160W image (Supplementary Fig.13).  The three slices are chosen as the best-fit major axis and its 1$\sigma$ upper and lower limits (red solid and dashed lines in Supplementary Fig.13). Each datapoint along the slice is an average of 4 pixels ($0.\!\!^{\prime\prime}24$) in width, i.e., about one image resolution element  ($0.\!\!^{\prime\prime}26$).   We stop the measurements when the data are indistinguishable from the 1$\sigma$ fluctuation of the background noise.  A total size of the galaxy is defined by the boundary where the 1D SB drops to the 1$\sigma$ background noise level. We estimate \obj's total size  to be ${\rm R}_{tot}=15\pm1$ kpc in radius, with the error bar indicating the uncertainty in identifying the boundary that is consistent with the noise.  The ${\rm R}_{tot}$ agrees with our measured ${\rm R}_{95} =15.6\pm0.6$ kpc using the circular aperture on the combined F125W+F140W+F160W  image (Supplementary Table 4).  We measure the ring's inner and outer radius (R$_{in}$ and R$_{out}$) based on the FWHM of the ring feature.  We use the  SB in the centre of the ring as the baseline of the FWHM. We find R$_{in}${=}2.1 kpc and R$_{out}${=}6.7 kpc,  broadly consistent with  the 2D double ellipse fit  (Supplementary Table 1).

The method we use to derive ${\rm R}_{tot}$ and ${\rm R}_{95}$  may not be practical for local CRGs such as the Cartwheel galaxy, where the ring dominates the luminosity in the outer disk (e.g., Supplementary Fig.14).  The method we use for ${\rm R}_{tot}$  is very similar to the commonly used ${\rm R}_{B25}$  for local CRGs\cite{Romano08}. ${\rm R}_{B25}$ refers to  the radius where the SB drops to a standard level of surface brightness of 25.00 mag arcsec$^{-2}$ in the B band for  angular dimensions\cite{deV91}.  Instead of using a fixed SB, we use the 1$\sigma$ sky background that is more suitable for high-redshift observations.

The average SB of the diffuse disk estimated from the average light between R$_{out}$ and R$_{95}$  is 0.42 $\mu$Jy arcsec$^{-2}$ in F160W. Assuming a cosmological SB dimming form of $(1+z)^{-4}$, the average SB of \obj's diffuse disk observed at $z\sim0$ would be 19.8 mag~arcsec$^{-2}$ in the B band.  This is almost four magnitudes brighter than the SB of the brightest outer disk of nearby CRGs\cite{Romano08}. Using the average SB of the diffuse disk as the baseline of the FWHM yields a $\sim$0.5 kpc increase/decrease in the size of the inner/outer radius.  The average SB of the ring as calculated between R$_{in}$ and R$_{out}$  is 1.04 $\mu$Jy arcsec$^{-2}$. The peak SB inside the ring is 1.15 $\mu$Jy arcsec$^{-2}$.  Therefore the relative SB  between the ring and the outer disk is 0.6-0.7 $\mu$Jy arcsec$^{-2}$.

\subsection{Comparing with 3D-HST $z\approx2$ late-type galaxies}
To put  the size of \obj\ in context with other $z\approx2$ galaxies, we compare its R$_{80}$ with  3D-HST galaxies measured on the same F160W images\citep{vander14} in Fig.3.  The sizes of 3D-HST galaxies have been modelled by Sersic profiles through  several well-established studies\cite{vander14,Mowla19, Miller19}.  We use the mass-size catalogue data from the 3D-HST survey\cite{vander14} and apply  the same conversion between  Sersic index $n$, effective radius R$_{e}$ and R$_{80}$ as previous studies\cite{vander14,Mowla19, Miller19}.  In order to minimise systematic errors of R$_{80}$, we only include galaxies with  flux SNR$>20$ on the F160W image and have  good Sersic model fits (SNR$>$5 for both R$_{e}$ and $n$).  We  select data with 3D-HST photometric redshifts ($zp$) of $1.8<z<2.4$ and $\log({\rm M_*}/{\rm M}_{\odot})>9.5$, to match our \zfourge\ target selection criteria. Only late-type  ($n<2.5$) galaxies are used.  We then cross-correlate the 3D-HST $zp$ with the high-precision ($\sim$2\%) \zfourge\ photometric redshifts\cite{Straatman16} and  exclude targets with $zp < 1.5$ or $zp  > 2.5$.  We also exclude targets that have inconsistent  stellar masses ($\Delta \log({\rm M_*}/{\rm M}_{\odot}) > 0.5$) from these two catalogues. Targets without \zfourge\  $zp$ or ${\rm M}_*$ remain in the sample. A total of $N=440$ objects satisfy  these selection criteria.

The empirical conversion between R$_{80}$,  R$_{e}$, and $n$ is based on large samples. The conversion is not guaranteed for individual galaxies. For example,  the R$_{80}$ for G4 and G5 in  Fig.3 is probably inaccurate and reflects the scatter in the empirical conversion.  \obj\ was included in the Sersic profile modelling of previous 3D-HST studies\cite{vander14}.  The inferred R$_{80}$ of \obj\ from the  3D-HST catelogue\cite{vander14} Sersic  R$_{e}$ (7.6 kpc) and $n$ (0.42)  is 11.2 kpc, in broad agreement with our non-parametric measurement  (11.8$\pm$0.3 kpc). Our model-independent R$_{50}$ for \obj\ is also consistent with the Sersic model-based R$_{e}$ from the 3D-HST catalogue\cite{vander14} within the measurement errors.  Both the R$_{80}$ and R$_{50}$ of  \obj\ are 1.5$\sigma$ above the scatter of the late-type galaxies at $z\approx2$ of similar or higher stellar masses ($\log({\rm M_*}/{\rm M}_{\odot})\ge10.6$).  We conclude that  \obj\ is an unusually large galaxy at $z\approx2$ regardless of the methods we use to quantify its total size.

The large size of \obj\ can be further appreciated when compared with the Milky Way. The scale length (R$_{s}$) of the Milky  Way's stellar disk is  in the range of 2-3 kpc based on a large body of literature\cite{Juric08,Wegg15,Bland16}.  Using R$_{50}${=}1.68 R$_{s}$\cite{Graham05,Glazebrook13r} and  R$_{80}\sim$3.2 R$_{s}$\cite{Giovanelli13}, our Milky Way has R$_{50}${=}3.4-5.0 kpc and R$_{80}${=}6.4-9.6 kpc in its stellar light.  According to our best size estimation from HST WFC3 images (Table 1),  the R$_{50}$ of \obj\ is  1.5-2.2 times larger and R$_{80}$ is 1.2-1.8 times larger than the Milky Way.

\section{A Joint  Analysis  of  MOSFIRE and OSIRIS \ha\ Kinematics}\label{sec:vel}
The line-of-sight velocity ($V_{\rm{LOS}}$) as a function of de-projected position angle ($\psi$) at a fixed radius on the ring is one of the commonly used methods to derive the  expansion/contraction and rotational velocities of CRGs.  Following similar  analysis of local CRGs\cite{Few82,Higdon96,Higdon11,Fogarty11}, we use  equations below to describe the relation between  $V_{\rm{LOS}}$  and $\psi$ in  a tilted rotating and expanding/contracting circular ring model. An illustration of the geometry and  definition of parameters is presented in Supplementary Fig.5: 
\begin{equation}\label{equ:eq1}
{\rm 
  V_{\rm LOS,\psi} =(V_{\rm sys} + (V_{\rm rad} ~cos (\psi) ~ - ~V_{\rm rot}  ~sin(\psi)))~sin(i), 
}
\end{equation}
\begin{equation}\label{equ:eq2}
{\rm 
  tan~(\psi) = tan(\psi_{o})~cos(i); ~  i \neq 90\degree.
}
\end{equation}
 $V_{\rm LOS,\psi}$  is the line-of-sight velocity measured at $\psi$ on the ring and $V_{\rm sys}$ is the systematic velocity.  $V_{\rm{LOS}}$ is calculated with respect to the kinematic centre of the galaxy. We take the cosmological expansion out by using the systematic redshift measured on the \ha\ centroid of the total aperture MOSFIRE spectrum ($z_{s,total}$),  hence we have  V$_{\rm sys}$ $\approx0$ if  $z_{s,total}$ is close to the systematic redshift  at the  kinematic centre.   A positive sign of  $V_{\rm{LOS}}$ means redshift  whereas a negative sign means blueshift. V$_{\rm rad}$ is the expansion/contraction velocity  at the fixed radius of the ring.   V$_{\rm rot}$ is the rotation velocity.   The angle $\psi$ is measured counterclockwise from the  kinematic major axis. The relation between the observed position angle ($\psi_{o}$) and the  deprojected $\psi$ is a simple function of the inclination angle $i$, where $i=0$ means viewing the disk of the ring face-on and $i=90$ means edge-on.  The corrections between $\psi_{o}$ and $\psi$ are small for face-on disks, whereas the equations are  invalid for edge-on disks (e.g., Supplementary Fig.5).   

Note that in  Eq.~\ref{equ:eq1},  the signs of V$_{\rm rad}$  and  V$_{\rm rot}$  can be either positive or negative.  In a fiducial  case in Supplementary Fig.5, we  define an east and a west side similar to the compass on a 2D image. 
In this case, the east side (with respect to the kinematic major axis) is the far side of the disk relative to the observer, a positive/negative sign of V$_{\rm rad}$ means expanding/contracting radially,  and a  positive/negative  sign of   V$_{\rm rot}$  means rotating counterclockwise/clockwise  from the  kinematic major axis.   The signs of V$_{\rm rad}$ and  V$_{\rm rot}$ are flipped if  the  east side is the near side of the ring. Without  knowledge of the near and far side of the ring, the  equations above provide only a magnitude of the radial velocity (V$_{\rm rad}$) and rotational velocity (V$_{\rm rot}$).  In the nearby universe, the near and far side of the ring can be determined from an extinction-reddening asymmetry across the minor axis in a tilted galactic disk.  This asymmetry is most evident when a galaxy has a prominent bulge: the bulge is viewed through the dust layer on the near side, while the dust is viewed through the bulge on the far side. 
 However, this method is difficult to apply to high-redshift galaxies whose bulge and disk are in the early stages of formation. 
   
The relation between  $V_{\rm{LOS}}$ and $\psi$ at a fixed radius (R) is therefore a function of four parameters:  the systematic velocity (V$_{\rm sys}$), the circular rotational velocity ($V_{\rm{rot}}$), the expansion/contraction velocity ($V_{\rm{rad}}$) and the inclination angle $i$ (Supplementary Fig.5). One additional  hidden parameter is the position angle (PA$_{0}$) of the kinematic major axis because $\psi$ is measured with respect to PA$_{0}$.   We do not have direct measurement of PA$_{0}$  and make the assumption that PA$_{0}$ is the position angle of the geometric major axis. We discuss the  consequence  of this assumption in the SI.
 
We can obtain three measurements on the  $V_{\rm{LOS}}$ versus $\psi$ diagram by combining our MOSFIRE and OSIRIS \ha\ kinematics.
The first two measurements are derived from the MOSFIRE slit spectrum and the third measurement comes from the OSIRIS observation.
Our MOSFIRE \ha\ velocity is spatially resolved and a clear relative wavelength separation is seen in the +y and -y spatial direction
 of the spectrum (Supplementary Fig.6).  We  align the MOSFIRE \ha\ 2D spectra in its spatial y direction with the NB209 \ha\ narrow-band image. We do this by 
 cross-correlating the spatial \ha\ line profiles of  the MOSFIRE observation with a mock \ha\ spatial line profile derived from the  \ha\ narrow-band image.
 In Supplementary Fig.6, we show the MOSFIRE spectrum aligned spatially  with the NB209 image after correcting for the central offset;  the error bar ($\sim0.\!\!^{\prime\prime}2$) of  the alignment is estimated by the 1$\sigma$ scatter of  100 times refit to the convoluted  NB209 image using a  full range of seeing ($0.\!\!^{\prime\prime}6$-$0.\!\!^{\prime\prime}9$) sizes experienced in our MOSFIRE observations. We then align the OSIRIS datacube with the NB209 \ha\ narrow band image using the  corrected astrometry (see Supplementary Information).  The maximum error of the OSIRIS to NB209 alignment is the spatial resolution of our adaptive optics (AO) observations ($0.\!\!^{\prime\prime}3$).

Our measurement is then carried out at the ring radius (R{=}5 kpc) for 3 positions  on the ring. We assume that the best-fit ellipse PA is the kinematic position angle PA$_{0}$ and  fix the inclination angle at the best value of 29$\degree$ as  derived from the ellipse fitting.  The uncertainties in $\psi$  are taken as the edge of the MOSFIRE slit and the edge of the OSIRIS \ha\ detection box  (Supplementary Fig.6). The line-of-sight velocities from the MOSFIRE spectra are based on the \ha\ emission line centroid of the 1D spectra extracted at the 5 kpc location (Fig.3). Due to the seeing-limited nature of the MOSFIRE observation,  the 1D spectra  extracted at the  5 kpc position  ($\pm1$ pixel) on MOSFIRE  is  prone to the uncertainty of the PSF characterisation.   We test the uncertainties of Vm1 and Vm2 by shifting the extracting centre of the  1D spectra within the   range of the spatial alignment error ($\sim0.\!\!^{\prime\prime}2$). Because of the strong correlation of the  spatial position  and  the split of the blue- and redshift of the 2D spectrum,   we find that the  Gaussian width of the line profiles  provides a conservative estimate of the measurement uncertainty in Vm1 and Vm2.   As long as the kinematic field is relatively face-on ($<$ 45\degree), the effect of beam-smearing in deriving the \ha\ line centroid is budgeted into the uncertainty from the line width.   The low SNR of our OSIRIS spectrum  prompts us to use the line width as an upper limit for the uncertainty in determining the line centroid.  The uncertainties of the  line-of-sight velocities along the MOSFIRE PA  (Vm1, Vm2) and OSIRIS \ha\  (Vo) are  therefore taken  as the Gaussian width of the line profiles. We also note that whether the measurement is done on a 5 kpc radius circle or the best-fit ellipse  does not change the result.

We  fitted  our three data points with Eq.~\ref{equ:eq1}-\ref{equ:eq2} using a $\chi^2$ minimisation procedure, weighted by the inverse square of the total error $\sigma_{x, y}$ {=} sqrt ($\sigma^2$$_{(V_{\rm LOS,\psi})}$ + $\sigma^2$$_{(\psi)}$). We keep V$_{\rm sys}$ at 0, though allowing it as a free parameter to account for any offsets  between  $z_{s,total}$ and  the redshift from the actual kinematic centre of the disk does not   change our main conclusion.  For our fixed  PA$_{0}=28\degree$ and $i=29\degree$, the best-fit expansion/contraction velocity is V$_{\rm{rad}}$ {=} 226 $\pm$ 90 km/s and V$_{\rm{rot}}$ {=} 90 $\pm$ 75 km/s (Fig.3).  The systematic errors of our kinematic measurement are discussed in the Supplementary Information. Nearby CRGs show a range of   V$_{\rm{rad}}$  (50-220 km/s) and V$_{\rm{rot}}$  (50-350 km/s). For most nearby CRGs, V$_{\rm{rot}}$   is larger than  V$_{\rm{rad}}$\cite{Few82, Higdon15},   though in some cases (e.g., Arp 147),  V$_{\rm{rot}}$  can be a few times smaller than  V$_{\rm{rad}}$\cite{Fogarty11}.

\section {Colour, Stellar Mass and Age}\label{sec:smass}
\subsection{Global} \label{sec:smass_glo}
We estimate the total stellar mass and  average age of the stellar population via spectral energy distribution (SED) fitting  to \zfourge\ multi-band photometry.  The total photometry is measured on  PSF matched \zfourge\ images.

The  publicly available catalogue of \zfourge\ provides stellar masses  based on the SED-fitting code \fast\cite{Kriek09b} in combination with the photometric redshift from EAzY\cite{Brammer08}. \fast\ determines the best-fit parameters of  SED models through a $\chi^{2}$ minimisation procedure\citet{Tomczak16}.  We use  36 passbands of \zfourge's total photometry based on  the stellar population synthesis (SPS) model grids\cite{BC03}, a Chabrier IMF\citet{Chabrier03}, a fixed solar metallicity (1.0 Z$_{\odot}$), an exponentially declining star formation history (SFH), and a Calzetti\citet{Calzetti00} extinction law.  The   \zfourge\ FAST output catalogue\cite{Tomczak16}  records a total stellar mass of $\log({\rm M_*}/{\rm M}_{\odot})$ {=} $10.78 \pm 0.03$ for \obj,  10.40 $\pm$ 0.04  for G5593, and $9.89 \pm 0.09$ for G5475.  We test the systematic uncertainties  of ${\rm M}_*$ against different SED fitting packages and find a systematic error of $\sim$0.2 dex (see in Supplementary Information).

To obtain a lower and upper limit for the stellar mass of \obj\, we  run FAST  assuming two extreme SFH:  a constant star-formation model (CSF) with emission lines and a 10 Myr truncated burst with no star formation afterwards (Supplementary Fig.10).  The  CSF provides  an upper limit to the stellar mass and age of \obj:  $\log({\rm M_*}/{\rm M}_{\odot})$ $< 10.8$ and t$_{age}$$ <$ 2 Gyr. The truncated star formation model provides a lower limit of  $\log({\rm M_*}/{\rm M}_{\odot})$ $> 10.5$ and  t$_{age}$$ >$ 50 Myr. These are the best constraints we can provide for \obj\ from current SED analysis. 

\subsection{Spatially resolved photometry} \label{sec:smass_spa}
In  \zfourge, a super deep K-band detection image (Ks$_{\rm tot}$) was created by combining FourStar/Ks-band observations with pre-existing K-band images\cite{Straatman16}.  The  Ks$_{\rm tot}$  image reveals an off-centre region that we speculate could be the ``nucleus" of the pre-collisional galaxy (Supplementary Fig.2).   We measure the aperture photometry of this postulated ``nucleus" region  and compare it with  regions on the ring.  We use an aperture of a diameter $0.\!\!^{\prime\prime}47$, corresponding to the  PSF size of the Ks$_{\rm tot}$  image.  We  then derive the colour differences of the nucleus and the ring based  on the PSF matched Ks$_{\rm tot}$ and 
  HST F125W, F160W images.   
 
 We find that the  nucleus  is 0.29 mag  redder (3.6$\sigma$ significance) than the average colour of the ring in F160W-Ks$_{\rm tot}$,  and 0.42 mag  redder (3.9$\sigma$ significance) in F125W-F160W.   
 The intrinsic colour difference might be larger without the beam-smearing of the PSF.  For example, using the HST original images (i.e., without convolving with the PSF of the Ks$_{\rm tot}$ band),  the  nucleus is  0.54 mag redder (5.0$\sigma$ significance) than the ring in F125W-F160W. The  redder colour of the ``nucleus'' region is consistent with the existence of an  off-centre nucleus commonly seen in CRGs\cite{Appleton96}. Well-known examples of CRGs with an off-centre nucleus are Arp 147 and NGC 985. Future high-resolution images in the optical and near-infrared bands are required to further confirm the location of this nucleus.

\section {Star Formation Rate}\label{sec:sfr}
\subsection{Total star formation rate}
The SFRs for \zfourge\ sources\citet{Tomczak16}  are based on the combined rest-frame infrared luminosity (L$_{\rm{IR, 8-1000 \mu m}}$) and rest-frame UV luminosity (L$_{\rm{UV, 1216-3000\angstrom}}$).   The UV+IR approach assumes that the IR emission of galaxies comes from dust heated by the UV light of massive stars\cite{Tomczak16}.  L$_{\rm{IR}}$ is measured on an IR spectral template fitted to the 24, 100 and 160 $\mu$m far-IR photometry.  The  far-IR photometry is measured using apertures of 3$^{\prime\prime}$-6$^{\prime\prime}$ from the MIPS/PACS imaging with the de-blending technique\citet{Labbe06}.  L$_{\rm{UV}}$ is measured on the EAzY photometric models\cite{Tomczak16}.   The SFR$_{\rm{UV+IR}}$ for \obj\ is calculated to be  80$\pm$0.2 \msun~yr$^{-1}$.  Note that because of the proximity of the neighbouring galaxy G5593 and  the large PSF (FWHM$>$4$^{\prime\prime}$) of the MIPS and PACS images, the main uncertainties in the SFR$_{\rm{UV+IR}}$ of  \obj\ and  G5593 come from the systematics of  de-blending  the two sources.  G5593 is  at a similar redshift as \obj\ and is a bright non-AGN source detected in the 24 $\mu$m with SFR$_{\rm{UV+IR}}$ {=} 123 $\pm$ 2 \msun~yr$^{-1}$. 
G5475 is a quiescent galaxy, consistent with its early-type morphology.
 
For comparison, we  determine  the dust-uncorrected SFR from the total \ha\ flux of the MOSFIRE slit spectrum and have SFR$_{\rm H\alpha(slit; with~dust)}$ 
{=} 3.7 $\pm$ 0.1 $M_{\odot}$ yr$^{-1}$.  For dust attenuation correction on \ha\  (A(\ha)), we infer from the stellar dust attenuation obtained via SED fitting (A$_{\rm{v,star}}\approx1.1$). 
 We test two methods on dust correction. We first use  the empirical relation of A(\ha)) and A$_{\rm{v,star}}$ for $z\approx2$ star-forming galaxies as a function of SFR(SED) and $M_*$ (SED)\citet{Reddy15}; 
 we have A(\ha)){=}0.21+A$_{\rm{v,star}}$. The dust corrected SFR$_{\rm H\alpha(slit)}$ is therefore  12.4 $\pm$ 0.3 $M_{\odot}$ yr$^{-1}$ for the first method.
 For the second method, we use the classic nebular attenuation curve\citet{Cardelli89} with $R_{\rm{v}} = 3.1$ and have A(\ha){=}2.53 $\times$ \rm{E(B-V)}$_{\rm{HII}}$;
 Assuming \rm{E(B-V)}$_{\rm{star}}$  {=} 0.44 $\times$  \rm{E(B-V)}$_{\rm{HII}}$\cite{Calzetti00}, we have A(\ha) {=} 1.42 $\times$ A$_{\rm{v,star}}$ or  A(\ha) {=} 1.42 $\times$ A$_{\rm{v,star}}$\cite{Steidel14,Tran15}.  The dust corrected SFR$_{\rm H\alpha(slit)}$ is therefore  15.6$\pm$0.4 $M_{\odot}$ yr$^{-1}$ for the second method.   Finally, we estimate the  slit loss factor by   
aligning the MOSFIRE slit on the PSF matched  \ha\ narrow-band image  and calculating the fraction of flux that is outside of the slit on the \ha\ narrow-band image.
We derive a slit-loss factor of 3.1$\pm$1.1.  The final slit-loss  and dust attenuation corrected  SFR$_{\rm H\alpha}$  is 38.4$\pm$13.6 $M_{\odot}$ yr$^{-1}$ for the first method and
48.4$\pm$17.2  $M_{\odot}$ yr$^{-1}$ for the second method, with  errors representing statistical errors contributed mainly by the slit-loss correction uncertainty.   
Though the \ha\ SFR is  $\sim$2$\times$  smaller than the  SFR$_{\rm{UV+IR}}$, the discrepancy is not surprising given the  large uncertainty in dust attenuation and slit loss.   

\subsection{Spatial distribution and $\Sigma$$_{\rm SFR}$}
Using the location of the best fit double ellipse and the \ha\ narrow band image, we estimate that $\sim$45-70\% of the star formation occurs on the ring. 
The upper limit is calculated by assuming  that  the \ha\ narrow band image traces all recent star formation activity. The lower limit is estimated by including 
only  \ha\ pixels within the ring that are more than 5$\sigma$ brighter than the average \ha\ surface brightness.  We caution that this estimation 
does not include uncertainties from the beam-smearing of the   \ha\ narrow band image, spatial variation of the dust attenuation, and the contribution of faint 
SF regions. 

Our current data do not have the spatial resolution and SNR to calculate the spatially resolved  $\Sigma$$_{\rm SFR}$. We provide two simple estimates for an average
 $\Sigma$$_{\rm SFR}$. We first use  SFR$_{\rm{UV+IR}}$ divided by the circular area within R$_{80}$ and have
  $\Sigma$$_{\rm SFR}$ $\sim0.2$~\msunyrkpc. We then use SFR$_{\rm{H}\alpha} = 38$~\msunyr\ divided by the area of the ring defined by the best-fit ellipse and have 
    $\Sigma$$_{\rm SFR}$ $\sim0.4$~\msunyrkpc.

\subsection{Comparing with local CRGs on the ${\rm M_*}$-SFR  relation}
The well-established correlation between SFR and stellar mass ${\rm M_*}$ is the so-called  ``main-sequence" for star-forming galaxies.
The slope and scatter of the ${\rm M_*}$-SFR  relation does not evolve significantly from $z\sim0$ to $z\approx2$ [ref.\cite{Pearson18}].  At at a fixed stellar mass, 
star-forming galaxies at $z\sim$2 have $\sim$20 times higher SFR compared with $z\sim0$ star-forming  galaxies. 
Based on SFR$_{\rm{UV+IR}}$,  \obj\ is a star-forming galaxy that lies within the 1$\sigma$ scatter ($\sim$0.3 dex) of the ${\rm M_*}$-SFR  relation at $z\approx2$ [ref.\cite{Tomczak16}].

Local CRGs have moderately higher SFR than other local spirals.  To compare \obj\ with local CRGs on the   ${\rm M_*}$-SFR  relation, 
we use a sample of local CRGs\cite{Romano08} that have both the SFR and  stellar mass  measured in a self-consistent way. We exclude
two CRGs  that have contaminations in their SFR from either a neighbouring galaxy or an AGN. 
We also include literature data for Arp 147 [ref.\cite{Fogarty11}] and the Cartwheel galaxy\cite{Crivellari09}, 
 as well as our Milky Way\cite{Licquia15}.  As there is no reported stellar mass for the Cartwheel galaxy, we scale 
 its dynamical mass to a stellar mass using the scaling relation of local galaxies\cite{Taylor10} (Supplementary Fig.11).

\begin{addendum}

\item[Data Availability]  The imaging data presented here are publicly available from the \zfourge\ survey website (\url{http://zfourge.tamu.edu/}) and
from the 3D-HST archive (\url{https://archive.stsci.edu/prepds/3d-hst/}). 
The spectroscopic data of this work was based on observations made with the Keck telescope from the W. M. Keck Observatory.
The raw spectroscopic data can be accessed through the  publicly available Keck Observatory Archive (\url{https://www2.keck.hawaii.edu/koa/public/koa.php}). 
The reduced data and other data that support the plots within this paper and other findings of this study are available  from the corresponding author on reasonable request.

\item[Code availability] 
The customised MOSFIRE spectroscopic fitting code used in this work can be found here
(\url{http://astronomy.swin.edu.au/~tyuan/mosfit/}).
Scripts related to EAGLE simulations analysis in this paper are available  from the second author  (A.E., email: ahmedagali70@gmail.com) on reasonable request.
Other scripts related to the analysis in this paper are available  from the corresponding author (T.Y.) on reasonable request.

\item[Supplementary information] accompanies this paper.  
The author's link to the Supplementary Information (SI) can be found here \url{http://astronomy.swin.edu.au/~tyuan/paper/}.  The  SI includes
10  sections, 16 figures and 4 tables.

\end{addendum}

\begin{flushleft}
\bf{\large References}
\end{flushleft}


\begin{thebibliography}{10}
\expandafter\ifx\csname url\endcsname\relax
\def\url#1{\texttt{#1}}\fi
\expandafter\ifx\csname urlprefix\endcsname\relax\def\urlprefix{URL }\fi
\providecommand{\bibinfo}[2]{#2}
\providecommand{\eprint}[2][]{\url{#2}}

\bibitem{Madore09}
\bibinfo{author}{Madore, B.~F.}, \bibinfo{author}{Nelson, E.} \& \bibinfo{author}{Petrillo, K.}
\newblock Atlas and Catalog of Collisional Ring Galaxies.
\newblock \emph{\bibinfo{journal}{\apjs}} \textbf{\bibinfo{volume}{181}},
 \bibinfo{pages}{572--604} (\bibinfo{year}{2009}).

\bibitem{Lynds76}
\bibinfo{author}{Lynds, R.} \& \bibinfo{author}{Toomrel, A.}
\newblock On the interpretation of ring galaxies: the binary ring system II Hz 4.
\newblock \emph{\bibinfo{journal}{\apj}} \textbf{\bibinfo{volume}{209}},
  \bibinfo{pages}{382--388} (\bibinfo{year}{1976}).

\bibitem{Struck93}
 \bibinfo{author}{Struck-Marcell, C.} \& \bibinfo{author}{Higdon, J.~L.}
\newblock Hydrodynamic models of the Cartwheel ring galaxy.
\newblock \emph{\bibinfo{journal}{\apj}} \textbf{\bibinfo{volume}{411}},
 \bibinfo{pages}{108-124} (\bibinfo{year}{1993}).


\bibitem{Appleton96}
\bibinfo{author}{Appleton, P.~N.}, \bibinfo{author}{Struck-Marcell, C.}
\newblock Collisional Ring Galaxies.
\newblock \emph{\bibinfo{journal}{FCPh}} \textbf{\bibinfo{volume}{16}},
  \bibinfo{pages}{111--220} (\bibinfo{year}{1996}).

\bibitem{Higdon95}
\bibinfo{author}{Higdon, J.~L.}
\newblock Wheels of Fire. I. Massive Star Formation in the Cartwheel Ring Galaxy.
\newblock \emph{\bibinfo{journal}{\apj}} \textbf{\bibinfo{volume}{455}},
 \bibinfo{pages}{524} (\bibinfo{year}{1995}).


\bibitem{Gerber96}
\bibinfo{author}{Gerber, R.~A.}, \bibinfo{author}{Lamb, S.~A.} \& \bibinfo{author}{Balsara, D.~S.}
\newblock A stellar and gas dynamical numerical model of ring galaxies. 
\newblock \emph{\bibinfo{journal}{\mnras}} \textbf{\bibinfo{volume}{278}},
  \bibinfo{pages}{345--366} (\bibinfo{year}{1996}).

 
\bibitem{Struck10}
 \bibinfo{author}{Struck, C.}
\newblock Applying the analytic theory of colliding ring galaxies.
\newblock \emph{\bibinfo{journal}{\mnras}} \textbf{\bibinfo{volume}{403}},
 \bibinfo{pages}{1516-1530} (\bibinfo{year}{2010}).


\bibitem{Mapelli12}
\bibinfo{author}{Mapelli, M.} \& \bibinfo{author}{Mayer, L.}
\newblock Ring galaxies from off-centre collisions.
\newblock \emph{\bibinfo{journal}{\mnras}} \textbf{\bibinfo{volume}{420}},
  \bibinfo{pages}{1158--1166} (\bibinfo{year}{2012}).

\bibitem{Higdon15}
\bibinfo{author}{Higdon, J.~L.}, \bibinfo{author}{Higdon, S.~J.~U.}, \bibinfo{author}{Mart{\'{\i}}n Ruiz, S.} \& \bibinfo{author}{Rand, R.~J.}
\newblock Molecular Gas and Star Formation in the Cartwheel.
\newblock \emph{\bibinfo{journal}{\apjl}} \textbf{\bibinfo{volume}{814}},
 \bibinfo{pages}{L1} (\bibinfo{year}{2015}).


\bibitem{Lavery04}
\bibinfo{author}{Lavery, R.~J.}, \bibinfo{author}{Remijan, A.}, \bibinfo{author}{Charmandaris, V.}, \bibinfo{author}{Hayes, R.~D.} \& \bibinfo{author}{Ring, A.~A.}
\newblock Probing the Evolution of the Galaxy Interaction/Merger Rate Using Collisional Ring Galaxies.
\newblock \emph{\bibinfo{journal}{\apj}} \textbf{\bibinfo{volume}{612}},
  \bibinfo{pages}{679--689} (\bibinfo{year}{2004}).

\bibitem{ElmegreenD06b}
\bibinfo{author}{Elmegreen, D.~M.} \& \bibinfo{author}{Elmegreen, B.~G.}
\newblock Rings and Bent Chain Galaxies in the GEMS and GOODS Fields.
\newblock \emph{\bibinfo{journal}{\apj}} \textbf{\bibinfo{volume}{651}},
  \bibinfo{pages}{676--687} (\bibinfo{year}{2006}).

\bibitem{DOnghia08}
\bibinfo{author}{D'Onghia, E.}, \bibinfo{author}{Mapelli, M.} \& \bibinfo{author}{Moore, B.}
\newblock Merger and ring galaxy formation rates at $z\le2$.
\newblock \emph{\bibinfo{journal}{\mnras}} \textbf{\bibinfo{volume}{389}},
  \bibinfo{pages}{1275--1283} (\bibinfo{year}{2008}).  


\bibitem{Elagali18}
\bibinfo{author}{Elagali, A.} \emph{et~al.}
\newblock Ring galaxies in the EAGLE hydrodynamical simulations.
\newblock \emph{\bibinfo{journal}{\mnras}} \textbf{\bibinfo{volume}{481}},
  \bibinfo{pages}{2951--2969} (\bibinfo{year}{2018}).


\bibitem{Genzel14}
\bibinfo{author}{Genzel, R.}  \emph{et~al.}
\newblock The SINS/zC-SINF Survey of $z\approx2$ Galaxy Kinematics: Evidence for Gravitational Quenching.
\newblock \emph{\bibinfo{journal}{\apj}} \textbf{\bibinfo{volume}{785}},
  \bibinfo{pages}{75} (\bibinfo{year}{2014}).


\bibitem{Buta96}
\bibinfo{author}{Buta, R.~J.}  \& \bibinfo{author}{Combes, F.}
\newblock Galactic Rings. 
\newblock \emph{\bibinfo{journal}{\fcp}} \textbf{\bibinfo{volume}{17}},
  \bibinfo{pages}{95--281} (\bibinfo{year}{1996}).

\bibitem{Comeron14}
\bibinfo{author}{Comeron, S.} \emph{et~al.}
\newblock ARRAKIS: atlas of resonance rings as known in the S$^{4}$G.
\newblock \emph{\bibinfo{journal}{\aap}} \textbf{\bibinfo{volume}{562}},
  \bibinfo{pages}{121} (\bibinfo{year}{2014}).



\bibitem{Sheth12}
\bibinfo{author}{Sheth, K.} \emph{et~al.}
\newblock  Hot Disks and Delayed Bar Formation.
\newblock \emph{\bibinfo{journal}{\apj}} \textbf{\bibinfo{volume}{758}},
 \bibinfo{pages}{136} (\bibinfo{year}{2012}).


\bibitem{Kraljic12}
\bibinfo{author}{Kraljic, F.}, \bibinfo{author}{Bournaud, F.} \&   \bibinfo{author}{Martig, M.}   
\newblock The Two-phase Formation History of Spiral Galaxies Traced by the Cosmic Evolution of the Bar Fraction.
\newblock \emph{\bibinfo{journal}{\apj}} \textbf{\bibinfo{volume}{757}},
 \bibinfo{pages}{60} (\bibinfo{year}{2012}).


\bibitem{Cen14}
\bibinfo{author}{Cen, R.}
\newblock Evolution of Cold Streams and the Emergence of the Hubble Sequence.
\newblock \emph{\bibinfo{journal}{\apjl}} \textbf{\bibinfo{volume}{789}},
  \bibinfo{pages}{L21} (\bibinfo{year}{2014}).


\bibitem{ElmegreenD14}
\bibinfo{author}{Elmegreen, D.~M.} \& \bibinfo{author}{Elmegreen, B.~G.}
\newblock The Onset of Spiral Structure in the Universe.
\newblock \emph{\bibinfo{journal}{\apj}} \textbf{\bibinfo{volume}{781}},
  \bibinfo{pages}{11} (\bibinfo{year}{2014}).


\bibitem{Vincenzo19}
\bibinfo{author}{Vincenzo, F.}, \bibinfo{author}{Kobayashi, C.}  \&  \bibinfo{author}{Yuan, T.}
\newblock Zoom-in cosmological hydrodynamical simulation of a star-forming barred, spiral galaxy at redshift $z=2$.
\newblock \emph{\bibinfo{journal}{\mnras}} \textbf{\bibinfo{volume}{488}},
 \bibinfo{pages}{4674-4689} (\bibinfo{year}{2019}).


\bibitem{Straatman16}
\bibinfo{author}{Straatman, C.~M.} \emph{et~al.}
\newblock The FourStar Galaxy Evolution Survey (ZFOURGE): 
Ultraviolet to Far-infrared Catalogs, Medium-bandwidth Photometric Redshifts with Improved Accuracy, Stellar Masses, and Confirmation of Quiescent Galaxies to $z\sim 3.5$.
\newblock \emph{\bibinfo{journal}{\apj}} \textbf{\bibinfo{volume}{830}},
  \bibinfo{pages}{51} (\bibinfo{year}{2016}).

\bibitem{Momcheva16}
\bibinfo{author}{Momcheva, I.~G.} \emph{et~al.}
\newblock The 3D-HST Survey: Hubble Space Telescope WFC3/G141 Grism Spectra, Redshifts, and Emission Line Measurements for $\sim$100,000 Galaxies.
\newblock \emph{\bibinfo{journal}{\apjs}} \textbf{\bibinfo{volume}{225}},
  \bibinfo{pages}{27} (\bibinfo{year}{2016}).


\bibitem{Cowley16}
\bibinfo{author}{Cowley, M.~J.} \emph{et~al.}
\newblock ZFOURGE catalogue of AGN candidates: an enhancement of 160-$\mu$m-derived star formation rates in active galaxies to $z = 3.2$.
\newblock \emph{\bibinfo{journal}{\mnras}} \textbf{\bibinfo{volume}{457}},
  \bibinfo{pages}{629--641} (\bibinfo{year}{2016}).


\bibitem{Romano08}
\bibinfo{author}{Romano, R.}, \bibinfo{author}{Mayya, Y.~D.} \& \bibinfo{author}{Vorobyov, E.~I.}
\newblock Stellar Disks of Collisional Ring Galaxies. I. New Multiband Images, Radial Intensity and Color Profiles, and Confrontation with N-Body Simulations.
\newblock \emph{\bibinfo{journal}{\aj}} \textbf{\bibinfo{volume}{136}},
  \bibinfo{pages}{1259--1289} (\bibinfo{year}{2008}).



\bibitem{Tacconi10}
\bibinfo{author}{Tacconi, L.~J.} \emph{et~al.}
\newblock High molecular gas fractions in normal massive star-forming galaxies in the young Universe.
\newblock \emph{\bibinfo{journal}{\nat}} \textbf{\bibinfo{volume}{463}},
  \bibinfo{pages}{781--784} (\bibinfo{year}{2010}).

  
\bibitem{Fogarty11}
\bibinfo{author}{Fogarty, L.} \emph{et~al.}
\newblock SWIFT observations of the Arp 147 ring galaxy system.
\newblock \emph{\bibinfo{journal}{\mnras}} \textbf{\bibinfo{volume}{417}},
 \bibinfo{pages}{853--844} (\bibinfo{year}{2011}).


\bibitem{Pearson18}
\bibinfo{author}{Pearson, W.~J.} \emph{et~al.}
\newblock Main sequence of star forming galaxies beyond the Herschel confusion limit.
\newblock \emph{\bibinfo{journal}{\aap}} \textbf{\bibinfo{volume}{615}},
  \bibinfo{pages}{A146} (\bibinfo{year}{2018}).



\bibitem{vander14}
\bibinfo{author}{van der Wel , A.}  \emph{et~al.}
\newblock 3D-HST+CANDELS: The Evolution of the Galaxy Size-Mass Distribution since $z = 3$.
\newblock \emph{\bibinfo{journal}{\apj}} \textbf{\bibinfo{volume}{788}},
  \bibinfo{pages}{28} (\bibinfo{year}{2014}).



\bibitem{Yuan17}
\bibinfo{author}{Yuan, T.-T.}  \emph{et~al.}
\newblock The Most Ancient Spiral Galaxy: A 2.6-Gyr-old Disk with a Tranquil Velocity Field.
\newblock \emph{\bibinfo{journal}{\apj}} \textbf{\bibinfo{volume}{850}},
  \bibinfo{pages}{61} (\bibinfo{year}{2017}).
  
\end{thebibliography}

\begin{thebibliography}{100}
\setcounter{enumiv}{30}


\expandafter\ifx\csname url\endcsname\relax
\def\url#1{\texttt{#1}}\fi
\expandafter\ifx\csname urlprefix\endcsname\relax\def\urlprefix{URL }\fi
\providecommand{\bibinfo}[2]{#2}
\providecommand{\eprint}[2][]{\url{#2}}


\bibitem{Planck14}
\bibinfo{author}{Planck Collaboration XVI} \emph{et~al.}
\newblock Planck 2013 results. XVI. Cosmological parameters.
\newblock \emph{\bibinfo{journal}{\aap}} \textbf{\bibinfo{volume}{571}},
 \bibinfo{pages}{16} (\bibinfo{year}{2014}).


\bibitem{Oke83}
\bibinfo{author}{Oke, J.~B.} \& \bibinfo{author}{Gunn, J.~E.}
\newblock Secondary standard stars for absolute spectrophotometry.
\newblock \emph{\bibinfo{journal}{\apj}} \textbf{\bibinfo{volume}{266}},
  \bibinfo{pages}{713--717} (\bibinfo{year}{1983}).


\bibitem{Grogin11}
\bibinfo{author}{Grogin, N.~A.}  \emph{et~al.}
\newblock CANDELS: The Cosmic Assembly Near-infrared Deep Extragalactic Legacy Survey.
\newblock \emph{\bibinfo{journal}{\apjs}} \textbf{\bibinfo{volume}{197}},
  \bibinfo{pages}{35} (\bibinfo{year}{2011}).


\bibitem{Skelton14}
\bibinfo{author}{Skelton, R.~E.}  \emph{et~al.}
\newblock 3D-HST WFC3-selected Photometric Catalogs in the Five CANDELS/3D-HST Fields: Photometry, Photometric Redshifts, and Stellar Masses.
\newblock \emph{\bibinfo{journal}{\apjs}} \textbf{\bibinfo{volume}{214}},
  \bibinfo{pages}{24} (\bibinfo{year}{2014}).

 
 \bibitem{Miller19}
\bibinfo{author}{Miller, T.~B.}  \bibinfo{author}{Gunn, J.~E.} \bibinfo{author}{van Dokkum, P.}  \bibinfo{author}{Mowla, L.} \& \bibinfo{author}{van der Wel, A.}
\newblock A New View of the Size-Mass Distribution of Galaxies: Using r$_{20}$ and r$_{80}$ Instead of r$_{50}$.
\newblock \emph{\bibinfo{journal}{\apjl}} \textbf{\bibinfo{volume}{872}},
  \bibinfo{pages}{L14} (\bibinfo{year}{2019}).
  

 \bibitem{Graham05}
\bibinfo{author}{Graham, A.~W.} \& \bibinfo{author}{Driver, S.~P.} 
\newblock A Concise Reference to (Projected) Sérsic R$^{1/n}$ Quantities, Including Concentration, Profile Slopes, Petrosian Indices, and Kron Magnitudes.
\newblock \emph{\bibinfo{journal}{Publications of the Astronomical Society of Australia}} \textbf{\bibinfo{volume}{22}},
  \bibinfo{pages}{118--127} (\bibinfo{year}{2005}).
 
  
  
  \bibitem{deV91}
\bibinfo{author}{de Vaucouleurs, G.}  \emph{et~al.}
\newblock Third Reference Catalogue of Bright Galaxies (RC3).
\newblock \emph{\bibinfo{journal}{Springer-Verlag, New York}} \textbf{\bibinfo{}{}},
  \bibinfo{}{} (\bibinfo{year}{1991}).
 
  
  
 \bibitem{Mowla19}
\bibinfo{author}{Mowla, L.}  \bibinfo{author}{van der Wel, A.} \bibinfo{author}{van Dokkum, P.} \& \bibinfo{author}{Miller, T.~B.}
\newblock A Mass-dependent Slope of the Galaxy Size-Mass Relation out to $z \sim 3$: 
Further Evidence for a Direct Relation between Median Galaxy Size and Median Halo Mass.
\newblock \emph{\bibinfo{journal}{\apjl}} \textbf{\bibinfo{volume}{872}},
  \bibinfo{pages}{L13} (\bibinfo{year}{2019}).
  
 \bibitem{Juric08}
\bibinfo{author}{Juric, M.} \emph{et~al.}
\newblock The Milky Way Tomography with SDSS. I. Stellar Number Density Distribution.
\newblock \emph{\bibinfo{journal}{\apj}} \textbf{\bibinfo{volume}{673}},
 \bibinfo{pages}{864--914} (\bibinfo{year}{2008}).

  
 \bibitem{Wegg15}
\bibinfo{author}{Wegg, C.}  \bibinfo{author}{Gerhard, O.} \& \bibinfo{author}{Portail, M.}
\newblock The structure of the Milky Way's bar outside the bulge.
\newblock \emph{\bibinfo{journal}{\mnras}} \textbf{\bibinfo{volume}{450}},
  \bibinfo{pages}{4050--4069} (\bibinfo{year}{2015}).
 

\bibitem{Bland16}
\bibinfo{author}{Bland-Hawthorn, J.} \& \bibinfo{author}{Gerhard, O.} 
\newblock The Galaxy in Context: Structural, Kinematic, and Integrated Properties.
\newblock \emph{\bibinfo{journal}{\araa}} \textbf{\bibinfo{volume}{54}},
  \bibinfo{pages}{529--596} (\bibinfo{year}{2016}).

 
 \bibitem{Glazebrook13r}
\bibinfo{author}{Glazebrook, K.} 
\newblock The Dawes Review 1: Kinematic Studies of Star-Forming Galaxies Across Cosmic Time.
\newblock \emph{\bibinfo{journal}{Publications of the Astronomical Society of Australia}} \textbf{\bibinfo{volume}{30}},
  \bibinfo{pages}{056} (\bibinfo{year}{2013}).
 
\bibitem{Giovanelli13}
\bibinfo{author}{Giovanelli, R.} 
\newblock On the scaling relations of disk galaxies.
\newblock \emph{\bibinfo{journal}{IAU Symposium}} \textbf{\bibinfo{volume}{289}},
  \bibinfo{pages}{296--303} (\bibinfo{year}{2013}).
 
 \bibitem{Few82}
 \bibinfo{author}{Few, M.~A.}, \bibinfo{author}{Madore, B.~F.}  \&  \bibinfo{author}{Arp, H.~C.}
 \newblock Ring galaxies. I - Kinematics of the southern ring galaxy AM 064-741.
 \newblock \emph{\bibinfo{journal}{\mnras}} \textbf{\bibinfo{volume}{199}},
  \bibinfo{pages}{633--647} (\bibinfo{year}{1982}).

\bibitem{Higdon96}
\bibinfo{author}{Higdon, J.~L.} 
\newblock Wheels of Fire. II. Neutral Hydrogen in the Cartwheel Ring Galaxy.
\newblock \emph{\bibinfo{journal}{\apj}} \textbf{\bibinfo{volume}{467}},
  \bibinfo{pages}{241} (\bibinfo{year}{1996}).
 
\bibitem{Higdon11}
\bibinfo{author}{Higdon, J.~L.}  \bibinfo{author}{Higdon, S.~J.~U.} \& \bibinfo{author}{Rand, R.~J.}
\newblock Wheels of Fire. IV. Star Formation and the Neutral Interstellar Medium in the Ring Galaxy AM0644-741.
\newblock \emph{\bibinfo{journal}{\apj}} \textbf{\bibinfo{volume}{739}},
  \bibinfo{pages}{97} (\bibinfo{year}{2011}).

 \bibitem{Kriek09b}
\bibinfo{author}{Kriek, M.} \emph{et~al.}
\newblock An Ultra-Deep Near-Infrared Spectrum of a Compact Quiescent Galaxy at $z = 2.2$. 
\newblock \emph{\bibinfo{journal}{\apj}} \textbf{\bibinfo{volume}{700}},
  \bibinfo{pages}{221--231} (\bibinfo{year}{2009}).
  
\bibitem{Brammer08}
\bibinfo{author}{Brammer, G.~B.}  \bibinfo{author}{van Dokkum, P.~G.} \& \bibinfo{author}{Coppi, P.}
\newblock EAZY: A Fast, Public Photometric Redshift Code. 
\newblock \emph{\bibinfo{journal}{\apj}} \textbf{\bibinfo{volume}{686}},
  \bibinfo{pages}{1503--1513} (\bibinfo{year}{2008}).

\bibitem{Tomczak16}
\bibinfo{author}{Tomczak, A.~R.} \emph{et~al.}
\newblock The SFR-M* Relation and Empirical Star-Formation Histories from ZFOURGE at $ 0.5 < z < 4 $.
\newblock \emph{\bibinfo{journal}{\apj}} \textbf{\bibinfo{volume}{817}},
  \bibinfo{pages}{118} (\bibinfo{year}{2016}).

 \bibitem{BC03}
\bibinfo{author}{Bruzual, G.} \& \bibinfo{author}{Charlot, S.} 
\newblock Stellar population synthesis at the resolution of 2003. 
\newblock \emph{\bibinfo{journal}{\mnras}} \textbf{\bibinfo{volume}{344}},
  \bibinfo{pages}{1000--1028} (\bibinfo{year}{2003}).
    
\bibitem{Chabrier03}
\bibinfo{author}{Chabrier, G.}
\newblock Galactic Stellar and Substellar Initial Mass Function. 
\newblock \emph{\bibinfo{journal}{\pasp}} \textbf{\bibinfo{volume}{115}},
  \bibinfo{pages}{763--795} (\bibinfo{year}{2003}).
  
 \bibitem{Calzetti00}
\bibinfo{author}{Calzetti, D.} \emph{et~al.}
\newblock The Dust Content and Opacity of Actively Star-forming Galaxies. 
\newblock \emph{\bibinfo{journal}{\apj}} \textbf{\bibinfo{volume}{533}},
  \bibinfo{pages}{682--695} (\bibinfo{year}{2000}).

 \bibitem{Labbe06}
\bibinfo{author}{Labbe, I.} \emph{et~al.}
\newblock Spitzer IRAC Confirmation of z850-Dropout Galaxies in the Hubble Ultra Deep Field: Stellar Masses and Ages at $z \sim 7$.
\newblock \emph{\bibinfo{journal}{\apjl}} \textbf{\bibinfo{volume}{649}},
  \bibinfo{pages}{L67-L70} (\bibinfo{year}{2006}).

\bibitem{Reddy15}
\bibinfo{author}{Reddy, N.~A.} \emph{et~al.}
\newblock The MOSDEF Survey: Measurements of Balmer Decrements and the Dust Attenuation Curve at Redshifts $z~1.4-2.6$.
\newblock \emph{\bibinfo{journal}{\apj}} \textbf{\bibinfo{volume}{806}},
  \bibinfo{pages}{259} (\bibinfo{year}{2015}).
  
\bibitem{Cardelli89}
\bibinfo{author}{Cardelli, J.~A.} \bibinfo{author}{Clayton, G.~C.} \& \bibinfo{author}{Mathis, J.~S.}
\newblock The relationship between infrared, optical, and ultraviolet extinction.
\newblock \emph{\bibinfo{journal}{\apj}} \textbf{\bibinfo{volume}{345}},
  \bibinfo{pages}{245--256} (\bibinfo{year}{1989}).
 
\bibitem{Steidel14}
\bibinfo{author}{Steidel, C.~C.} \emph{et~al.}
\newblock Strong Nebular Line Ratios in the Spectra of $z \sim 2-3$ Star Forming Galaxies: First Results from KBSS-MOSFIRE.
\newblock \emph{\bibinfo{journal}{\apj}} \textbf{\bibinfo{volume}{795}},
  \bibinfo{pages}{165} (\bibinfo{year}{2014}).

\bibitem{Tran15}
\bibinfo{author}{Tran, K.~V.~H.} \emph{et~al.}
\newblock ZFIRE: Galaxy Cluster Kinematics, H alpha Star Formation Rates, and Gas Phase Metallicities of XMM-LSS J02182-05102 at $zcl = 1.6232$.
\newblock \emph{\bibinfo{journal}{\apj}} \textbf{\bibinfo{volume}{811}},
  \bibinfo{pages}{28} (\bibinfo{year}{2015}).
  
\bibitem{Crivellari09}
\bibinfo{author}{Crivellari09, E.} ,\bibinfo{author}{Wolter, A.} \&  \bibinfo{author}{Trinchieri, G.}
\newblock The Cartwheel galaxy with XMM-Newton.
\newblock \emph{\bibinfo{journal}{\aap}} \textbf{\bibinfo{volume}{501}},
  \bibinfo{pages}{445--453} (\bibinfo{year}{2009}).

\bibitem{Licquia15}
\bibinfo{author}{Licquia, T.~C.}  \&  \bibinfo{author}{Newman, J.~A.}
\newblock Improved Estimates of the Milky Way's Stellar Mass and Star Formation Rate from Hierarchical Bayesian Meta-Analysis.
\newblock \emph{\bibinfo{journal}{\apj}} \textbf{\bibinfo{volume}{806}},
\bibinfo{pages}{96} (\bibinfo{year}{2015}).
    
\bibitem{Taylor10}
\bibinfo{author}{Taylor, E.~N.} \emph{et~al.}
\newblock On the Masses of Galaxies in the Local Universe.
\newblock \emph{\bibinfo{journal}{\apj}} \textbf{\bibinfo{volume}{722}},
\bibinfo{pages}{1-19} (\bibinfo{year}{2010}).


  
\end{thebibliography}
\end{document}